\documentclass[onecolumn]{aa}
\usepackage{graphicx,epsf,epsfig}

\usepackage{txfonts}
\newcommand{\Porb}{\mbox{$P_{\rm orb}$}}
\newcommand{\Pspin}{\mbox{$P_{\rm spin}$}}
\newcommand{\ecs}{\mbox{$\rm ergs\;cm^{-2}s^{-1}$}}
\newcommand{\es}{\mbox{$\rm ergs\;s^{-1}$}}

\begin{document}
   \title{X-ray Confirmation of the Intermediate Polar HT\,Cam}
\fnmsep\thanks{Based on observations obtained with XMM-Newton, an ESA
science mission with instruments and contributions directly funded by ESA 
Member States and NASA.}


   \author{D. de Martino
          \inst{1}
          \and
          G. Matt\inst{2}\and K. Mukai\inst{3}\and
J.-M. Bonnet-Bidaud\inst{4}, B.T. G\"ansicke\inst{5},
J.M. Gonzalez Perez\inst{6},
F. Haberl\inst{7}, M. Mouchet\inst{8}\and J.-E. Solheim \inst{9}
          }

\offprints{D. de Martino}

   \institute{INAF--Osservatorio Astronomico di Capodimonte, Via
Moiariello 16, I-80131 Napoli, Italy\\
              \email{demartino@na.astro.it}
         \and
Dipartimento di Fisica, Universita' degli Studi Roma Tre, Via della
Vasca Navale 84, I-00146 Roma, Italy \\
\email{matt@fis.uniroma3.it}
\and
Laboratory for High Energy Astrophysics, NASA/GSFC, Code 662, Greenbelt,
MD 20771, USA\\
 \email{mukai@milkyway.gsfc.nasa.gov}
\and
Service d'Astrophysique, DSM/DAPNIA/SAp, CE Saclay, F-91191 Gif sur Yvette
Cedex, France\\
\email{bonnetbidaud@cea.fr}
\and
Department of Physics, University of Warwick, Coventry CV4 7AL,
UK\\ 
\email{ boris.gaensicke@warwick.ac.uk}
\and
Instituto de Astrofisica de Canarias, Via Lactea, E-38205, La Laguna 
Tenerife, Spain \\
 \email{jgperez@iac.es}
\and
Max-Planck-Instit\"ut f\"ur Extraterrestrische Physik,
Giessenbachstra{\ss}e, Postfach 1312, 85741 Garching, Germany \\
\email{fwh@mpe.mpg.de}
\and
APC, UMR 7164, University Denis Diderot, 2 place Jussieu, F-75005 and 
LUTH, Observatoire de Paris, F-92195 Meudon Cedex, France\\
\email{martine.mouchet@obspm.fr} 
\and
Institute of Theoretical Astrophysics, University of Oslo, P.Box 1029,
N-0315 Blindern-Oslo\\
\email{janerik@phys.uit.no}
             }

   \date{Received 2005 01, 25; accepted 2005 03, 19}

   \abstract{
We report on  the first pointed X-ray observations with {\em 
XMM-Newton} 
and {\em RXTE} satellites of the X-ray source
RX\,J0757.0+6306=HT\,Cam. We detect a strong 515\,s X-ray modulation 
confirming the
optical photometric period found in 1998, which definitively assigns
this source to the intermediate polar class of magnetic cataclysmic
variables. The lack of orbital sidebands in the X-rays indicates that
the X-ray period is the spin period of the accreting white 
dwarf. Simultaneous ultraviolet
and optical B-band photometry acquired with the  XMM-Newton  Optical 
Monitor 
and
coordinated optical UBVRI photometric data acquired at the Nordic
Optical Telescope (La Palma) show that the optical pulse is in phase with
the X-rays and hence originates in the magnetically confined accretion 
flow. The lack of ultraviolet spin
modulation suggests that accretion-induced heating on the white dwarf
surface is not important in this source. Spectral analyses of  
XMM-Newton EPIC and RGS
spectra show that HT\,Cam has a multi-temperature spectrum and, contrary 
to most intermediate polars, 
it does not suffer from strong absorption. With its 86\,min orbital
period, HT\,Cam is then the third confirmed system of this 
class below 
the 2-3\,hr period gap accreting  at a low rate.

   }

   \maketitle
%

\section{Introduction}

RX\,J0757.0+6306=HT\,Cam (henceforth HT\,Cam) was identified  as a short
orbital
period (81\,min) cataclysmic variable (CV) 
and proposed as an intermediate polar (IP) by
Tovmassian et al. (\cite{Tovmassianetal})
for its rapid photometric variability at a period of 8.5\,min. 
These periods were recently refined from spectroscopy (\Porb=85.98\,min) 
and from extensive photometry (\Pspin=515.06\,s) by Kemp et al. 
(\cite{Kempetal}),  further 
supporting the IP identification. Whether the short period pulse 
represents the
spin period of the accreting white dwarf (WD) or an orbital sideband could
not be assessed with the optical photometry alone (see also Warner
(\cite{warner}) for the origin of optical pulsations in IPs).
In IPs the accretion flow  
is channelled by the relatively strong ($\rm B\leq$ 5-10\,MG) magnetic
field of the WD onto its polar regions where a
strong shock develops above the WD surface, below which the
gas cools via thermal bremsstrahlung peaking in the hard X-ray
regime. Hence, the signature of channelled accretion in these systems is
a strong X-ray pulsation at the WD rotational period which could
not be confirmed by the short ROSAT observation performed 
during the All  Sky Survey on this source.
HT\,Cam also occasionally displays short outbursts (Kemp et
al. \cite{Kempetal}, Ishioka et al. \cite{ishiokaetal}), with a decline 
rate of $\sim$ 4\,mag\,day$^{-1}$, thus sharing this peculiarity with 
another short orbital period IP, EX\,Hya.

   \begin{table*}[t!]
      \caption{Summary of the observations of HT\,Cam.}
         \label{obslog}
     \centering
\begin{tabular}{c c c c r c }
            \hline \hline
            \noalign{\smallskip}
Telescope   &  Instrument     &  Date & UT(start) & Duration (s)\\
            \noalign{\smallskip}
            \hline
            \noalign{\smallskip}
{\em XMM-Newton}    & EPIC PN &  2003 March 24 & 13:21 & 38997  \\
                      & EPIC MOS & & 12:58 & 39810 \\
                      & RGS    & & 12:58 & 40523      \\
                      & OM B & & 13:08 & 3499 \\
                      &      & & 14:11 & 3500 \\
                      &      & & 15:15 & 3498 \\
                      &      & & 16:18 & 3499 \\
                      &      & & 17:22 & 3499 \\
                      & OM UVW2 & & 18:26 & 3500 \\
                      &         & & 19:30 & 3500 \\
                      &         & & 20:34 & 3498 \\
                      &         & & 21:38 & 3499 \\
                      &         & & 22:41 & 3500 \\
                 \noalign{\smallskip}
            
{\em RXTE}    & PCA & 2002 Dec. 26 & 21:57 & 25000  \\
              &     & 2002 Dec. 27 & 21:51 & 24000 \\
  
                 \noalign{\smallskip}
{\em NOT}    &    ALFOSC V  & 2002 Dec. 26 & 20:55 & 12600  \\  
             & ALFOSC B  & 2002 Dec. 27 & 03:15 &  4440  \\  
             & ALFOSC U  & 2002 Dec. 27 & 20:25 & 11220   \\
             & ALFOSC R & 2002 Dec. 28 & 00:46 & 10920 \\  
             & ALFOSC I & 2002 Dec. 28 & 04:46 &  5820 \\  

                 \noalign{\smallskip}
            \hline
\end{tabular}
\end{table*}

 In the framework of a programme aiming at studying the X-ray properties
of new magnetic CVs, we were granted {\em XMM-Newton\/} and {\em RXTE\/}
observations during AO2 (PI:de Martino) and cycle 7
(PI: Mukai) respectively, which are described  in Sect.\, 2.
Furthermore, a coordinated optical multicolor photometric run at the 
Nordic Optical Telescope ( {\em NOT\/}) (La Palma) was carried out 
simultaneously
with the {\em RXTE\/} observation, also reported in Sect.\,2. 
We present the X-ray and optical/UV timing analyses in Sect.\,3 and 4
respectively and the X-ray spectral analysis is reported in Sect.\,5.  
We discuss the implications of X-ray and optical results in Sect.\,6 in
terms of accretion parameters and evolutionary state.


\section{The observations}

\subsection{The {\em XMM-Newton} observation}
  
HT\,Cam was observed with {\em XMM-Newton\/} satellite (Jansen et
al. \cite{jansenetal}) on 2003 March 24
(obsid:0144840101) with the EPIC-PN (Str\"uder et
al. \cite{strudeetal}) and MOS (Turner et al. \cite{turneretal}) cameras
operated in full frame mode   with the medium filter for  a net
exposure time of
39.0\,ks and 39.8\,ks respectively. HT\,Cam was also observed with the
Reflection Grating Spectrographs (RGS1 and RGS2) (den Herder et
al. \cite{denherderetal}) in
spectroscopy mode with an exposure time of
40.5\,ks and with the Optical Monitor (OM) instrument (Mason et
al. \cite{masonetal}) with the UVW2 and B filters
covering the ranges 1800--2250\,\AA \, and
3900--4900\,\AA \, in imaging fast mode for a total exposure time of
17.5\,ks in each filter. A summary of
the observations is reported in Table~\ref{obslog}.

The standard pipeline processing data products were used. 
The EPIC light curves and spectra and the RGS spectra were extracted with
the SAS 6.0 package retrieved  from the  ESA-VILSPA Science Center.
EPIC light curves and spectra of HT\,Cam were extracted
from circular regions with a radius of 40$^{"}$ centred on the source,
while background light curves and spectra were extracted from an offset
circular
region with radius of 90$^{"}$. Background subtracted EPIC net light
curves were obtained taking into account the different areas of 
extraction regions. Only single and double pixel events with a
zero quality flag were selected for the EPIC-PN data, while for EPIC-MOS
cameras all valid patterns have been used ( see 
{\em XMM-Newton\/} SAS User Guide:  
http://xmm.vilspa.esa.es/external/xmm$\_$user$\_$support/documentation/sas$\_$usg/USG).  
The EPIC-PN and, at a less extent, the MOS data, are  affected by
a strong background activity and hence data were windowed in order 
to exclude the high background epochs.  
The extracted EPIC-PN and MOS spectra were rebinned to have 
 a minimum of 25 counts in each bin to allow the use of the $\chi^2$ 
statistics.
Ancillary response and redistribution matrix files were created using SAS
meta-tasks {\em arfgen} and {\em rmfgen} respectively.
  
The RGS pipeline was run using the SAS meta-task {\em rgsproc}, and 
adopting
the same temporal windowing used for the EPIC cameras to exclude high 
background
intervals. RGS1 and RGS2 first order spectra have been rebinned to have 
a minimum of 20 counts per bin. Ancillary and response files were also
created for the RGS1 and RGS2 spectra.

OM background subtracted light curves, binned in 10\,s intervals, produced 
by the standard processing pipeline  were used for timing analysis.
  Average net count rates are 5.32\,cts\,s$^{-1}$ 
(B filter) and  0.326\,cts\,s$^{-1}$ (UVW2 filter). Using the zero
points available at the VILSPA  on-line documentation 
(http://xmm.vilspa.esa.es/external/xmm$\_$user$\_$support/documentation/uhb/node75.html, 
the instrumental magnitudes are B=17.2$\pm$0.5\,mag and
UVW2=16.1$\pm$0.7\,mag. Using Vega magnitude to flux conversion, these
correspond to a flux of $\rm 8.09\times 
10^{-16}\,erg\,cm^{-2}\,s^{-1}\,\AA^{-1}$
in the 3900--4900\,\AA \, band and of $\rm 2.04\times
10^{-15}\,erg\,cm^{-2}\,s^{-1}\,\AA^{-1}$
in the 1800--2250\,\AA \, band. The B band flux is
then similar to that observed during quiescence. 

\subsection{The {\em RXTE} data}

HT Cam was observed with  {\em RXTE}/PCA instrument on 2002 December
26  and again on December 27. Although the PCA comprises 5 proportional counter
units (PCUs), PCU0 suffers from a high background rate due to a leak
in the propane layer, and PCUs 1, 3, and 4 are often switched off for
instrumental reasons.  We have therefore analyzed only the PCU2 data.
After standard screening, good data were obtained for 25 ksec on
the first observation and 24 ksec on the second.  Background subtraction
was performed using the "Faint" background model, with calibration
file dated 2003 November 13.
HT\,Cam is detected with RXTE, however near the limit of background
subtraction uncertainties.  A spectral fit using the best-fit EPIC
model (cfr. Sect.\,5) suggests a similar flux level as during the  {\em
XMM-Newton} observations, although  a systematic pattern in the
residuals is present.

\subsection{The UBVRI photometry}

HT\,Cam was observed on 2002 December 26/27 and 27/28 at the {\em 
NOT} 2.56\,m telescope (La Palma) equipped with ALFOSC camera operated in 
windowed fast
photometric mode with UBVRI  filters simultaneously with 
the {\em 
RXTE} observations. Exposure times were such that the time
resolution of CCD data acquisition (exposure, read-out and dead times) was 
of 60\,s (U), 30\,s (B and V) and 20\,s
(R and I). The details of the observations are reported in  
 Table~\ref{obslog}. During the first night the photometric
conditions were not good due to strong wind especially during the
second part of the night, thus only a short coverage with the B 
band filter could be performed. The processing of the CCD images
was performed using {\em rtp} and {\em rtcorr} softwares available at the
{\em NOT} using aperture photometry, which include flat field correction,
sky subtraction, differential photometry
with several comparison stars in the field, extinction correction and data
normalisation by least square fitting. 
For all filters the same comparison star has been used to analyse the
light curves. 

   \begin{figure*}
   \centering
\includegraphics[height=13.cm,width=7.cm,angle=-90]{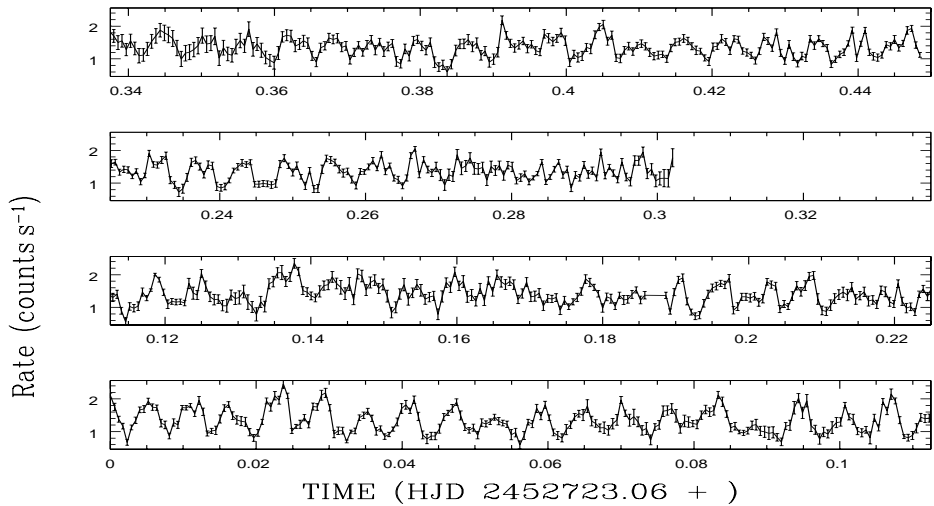}
\caption{EPIC-PN light curve in the 0.2--15\,keV range binned in 50\,s
intervals. High background epochs have been windowed as discussed in the
text}\label{fig1}
    \end{figure*}
   \begin{figure}
   \centering
\includegraphics[height=15.cm, width=14.cm,angle=-90]{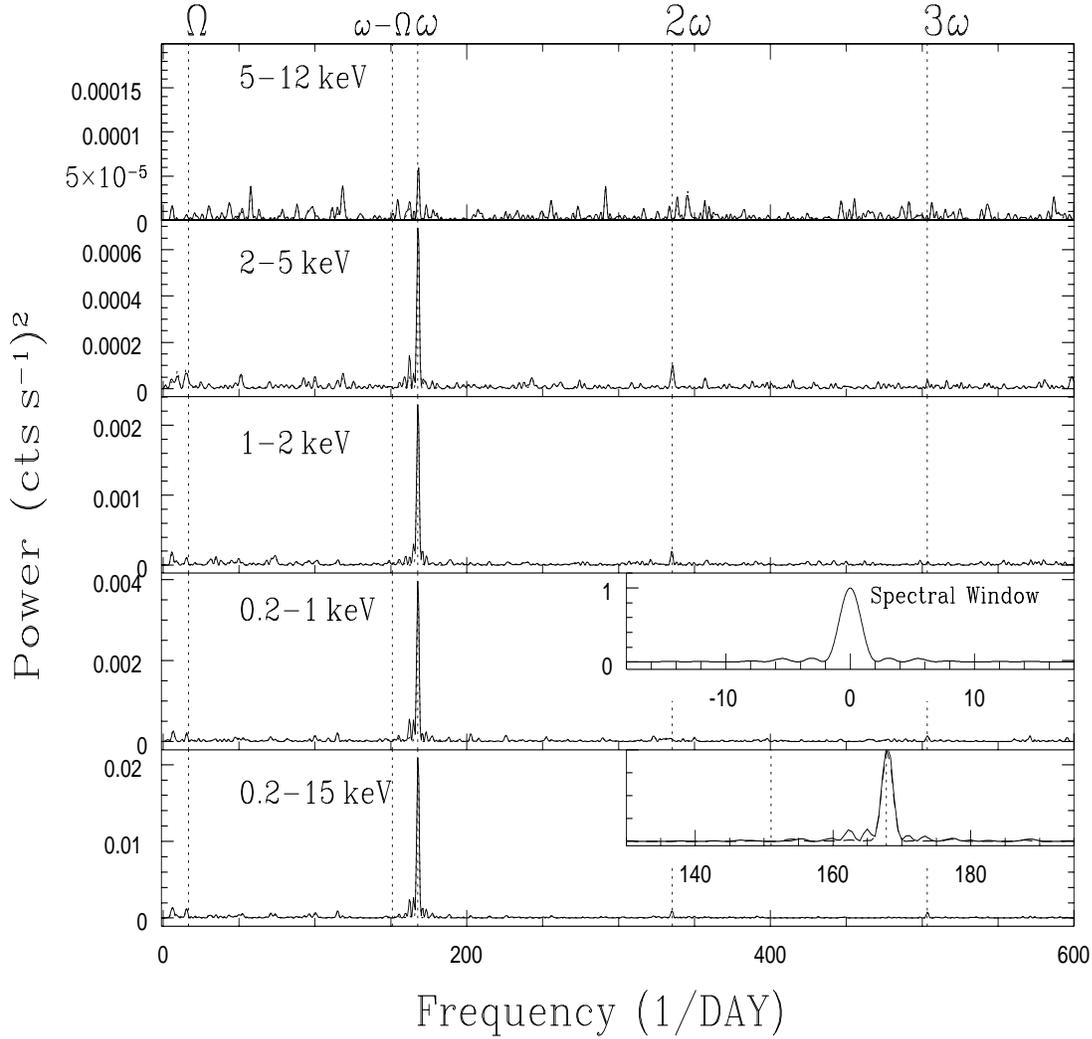}
\caption{EPIC-PN power spectra in selected energy ranges. From bottom 
to top: 0.2--15\,keV, 0.2--1\,keV, 1--2\,keV, 2--5\,keV and
5--12\,keV. The spin ($\omega$) and harmonics, the beat ($\omega -
\Omega$) and the orbital ($\Omega$) frequencies are marked with vertical
dotted lines. An enlargment around  the spin frequency is shown in 
the inserted 
bottom panel where the DFT (solid line) together with the CLEANED spectrum 
(dashed line) are shown. The spectral window is also shown.  }\label{fig2}
    \end{figure}

\section{X-ray Timing Analysis}

\subsection{The {\em XMM-Newton} light curves and power spectra}

 EPIC PN and MOS light curves in the 0.2--15\,keV range were
extracted with a 50\,s time resolution and in Fig.~\ref{fig1} we report
the
PN net light curve only as they all
show a si\-mi\-lar behaviour. A clear indication of a variability on time
scales of several minutes is found. The time series have been Fourier
analysed to detect periodic signals as shown in Fig.~\ref{fig2}. A strong
peak is
found at 168\,day$^{-1}$ with no other significant peak
close-by. The CLEAN algorithm (Roberts et al. \cite{robertsetal})
was also applied to remove windowing effects introduced by the
exclusions of high background epochs. This enables us to definitively
exclude the presence of orbital sidebands. Weak signals at the first and
second harmonic are also observed. No significant peak is found at
the orbital frequency. A sinusoidal fit to the PN and MOS light curves
gives $\omega$=167.90$\pm$0.05\,day$^{-1}$ for the PN,
168.04$\pm$0.07\,day$^{-1}$ for MOS1 and 167.88$\pm$0.08\,day$^{-1}$ for
MOS2. The inclusion of additional sinusoids, accounting for the first and
second harmonics does not improve the fit quality. We then adopt the EPIC 
PN period determination: $\rm P_{spin}$=514.59$\pm$0.15\,s  which
agrees within 3$\sigma$  with the refined optical pulse 
period found by Kemp et al. (\cite{Kempetal}). We hence determine 
the following ephemeris from  the  PN data for the time of maximum of 
X-ray spin
pulse: $\rm HJD_{max}$=2\,452723.28316(4)+0.005956(2)\,E. Here we 
note that the time of X-ray maximum is  consistent with that 
predicted by the accurate ephemeris by Kemp et al. (\cite{Kempetal}), 
which is derived from  data taken  between 1997 and 2002. 

\begin{figure*}[h]
   \centering
\mbox{\epsfxsize=8cm\epsfysize=8cm\epsfbox{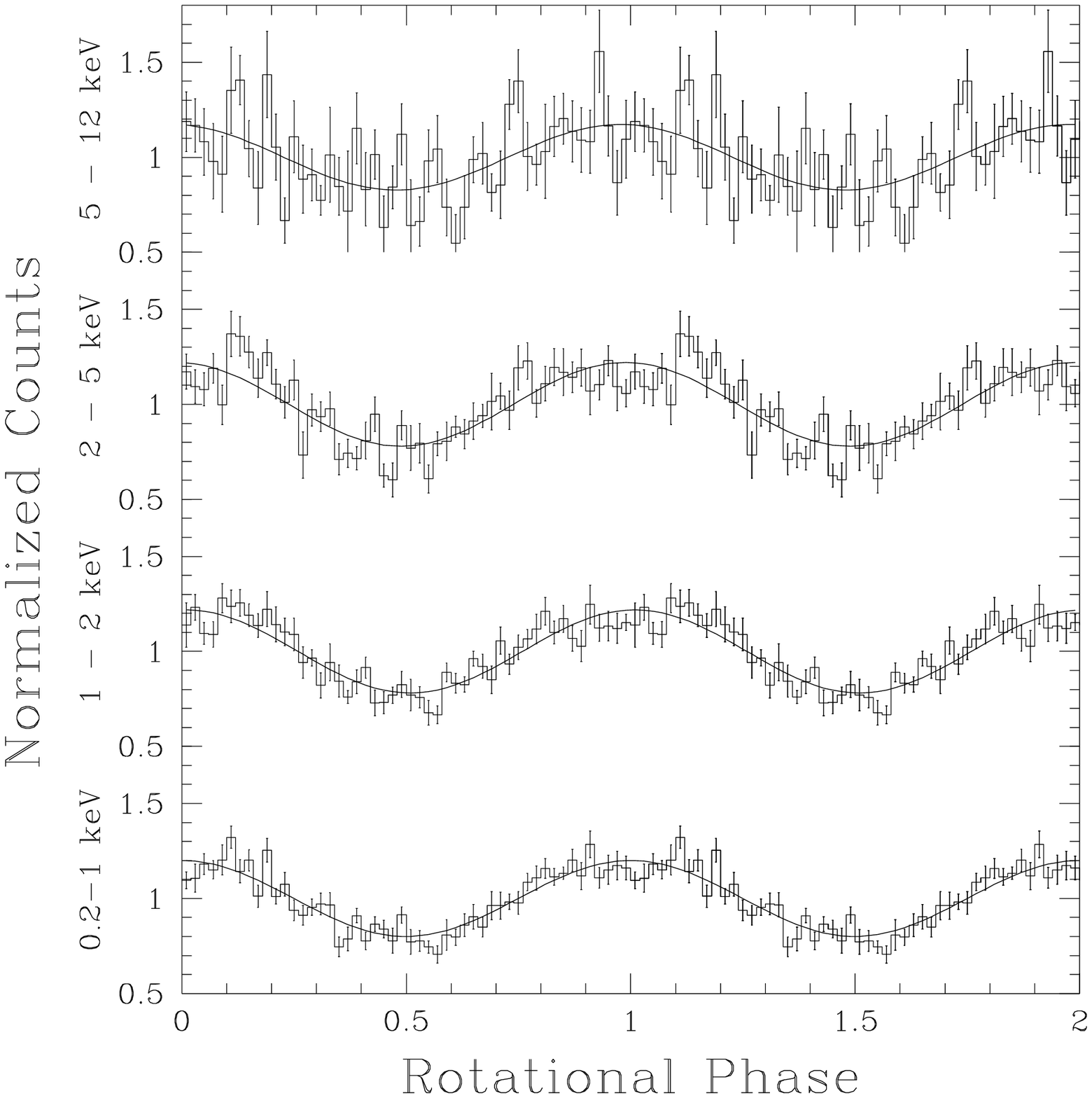}
\epsfxsize=8.cm\epsfysize=8cm\epsfbox{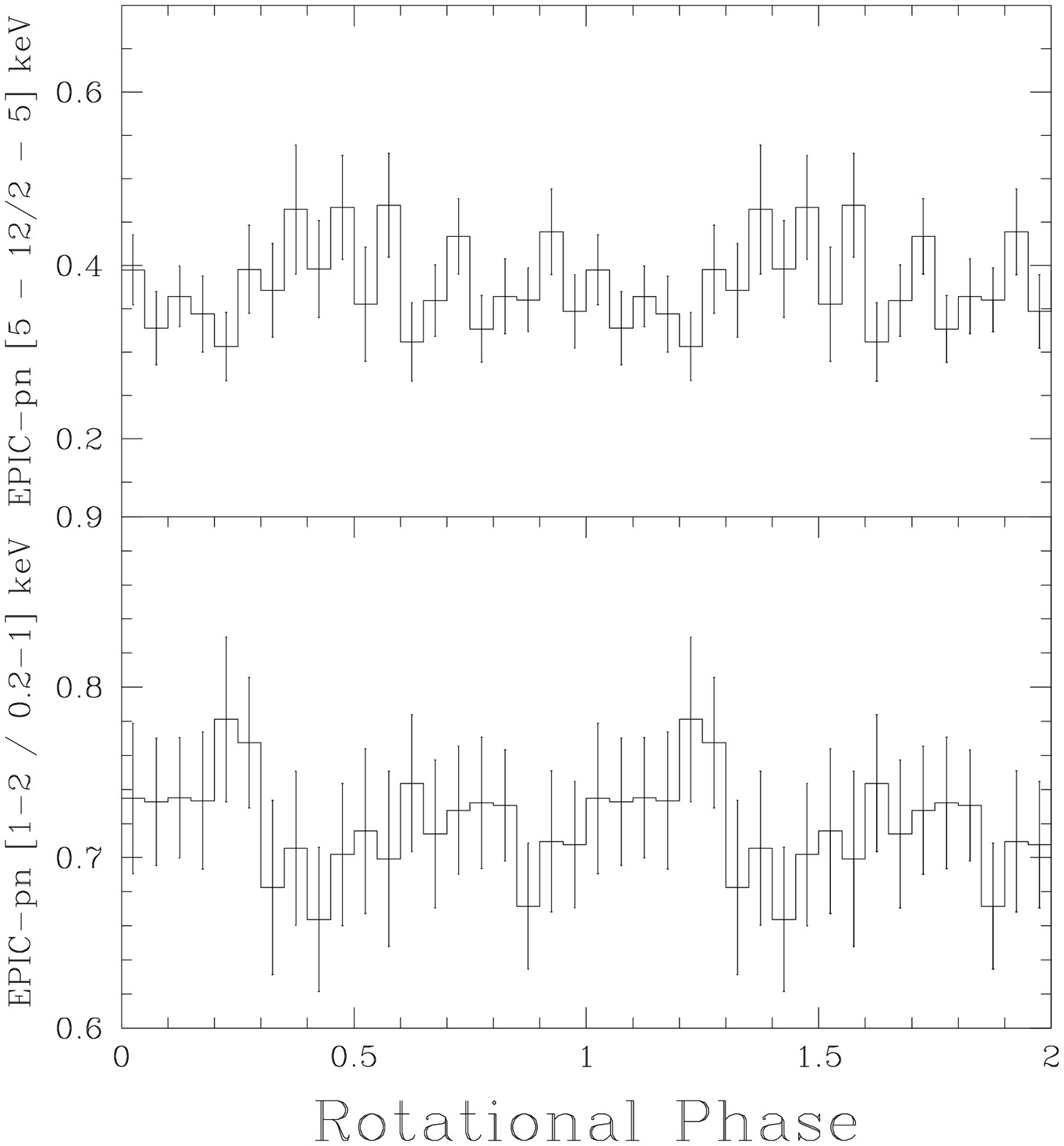}}
\caption{ {\em Left:} EPIC-PN folded light curves in selected energy bands
at the 515\,s period using the ephemeris quoted in the text together 
with the best sinusoidal fit. {\em Right:}
The EPIC hardness ratios show no energy dependence of the
pulse.}\label{fig3}
    \end{figure*}

\noindent The periodic variability was also explored  at different
energies in the EPIC range. Light curves were also extracted with the same
50\,s binning time in the four ranges: 0.2--1\,keV, 1--2\,keV, 2--5\,keV
and 5--12\,keV and Fourier analysed as shown in Fig.~\ref{fig2}. The
spin peak is strong at all energies except for the harder range where
it is detected at 2.8$\sigma$. The first and second harmonics are detected
in the 1-5\,keV range. 
Pulse folded light curves were  constructed
in the above ranges using the ephemeris given above. These are shown 
in Fig.~\ref{fig3}, together with the
hardness ratios defined as the ratio of countrates in [5--12\,keV] and
[2--5\,keV] ranges and in [1--2\,keV] and [0.2--1\,keV] ranges.
The light curves are quasi-sinusoidal and display  a broad maximum 
with an additional peak which is best visible in the 2--5\,keV band. 
The modulation has fractional amplitudes: $40 \pm 2\%$
in 0.2--1\,keV, $43 \pm 5\%$ in 1--2\,keV, 44$\pm 7\%$ in 2--5\,keV, 
and  34$\pm 15\%$ in 5--12\,keV band. Hence,  within the errors the 
pulsation has  the same amplitudes at all energies. 
The constancy of hardness ratios with spin phase further confirms the lack 
of energy dependence of pulse amplitudes.

\subsection{The RXTE light curve }
\begin{figure}[h]
\centering
\includegraphics[width=8cm]{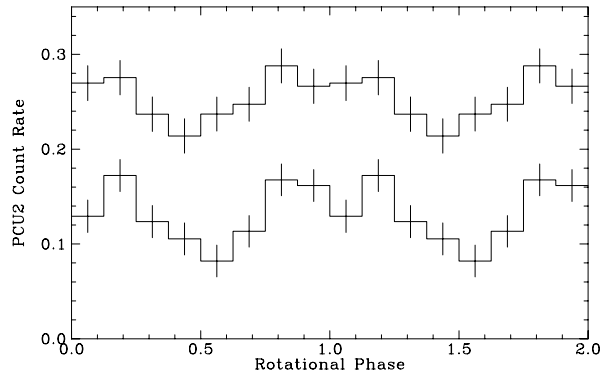}
      \caption{
The {\em RXTE} spin folded light curves in the 2--6\,keV 
(higher curve) and 6--10\,keV (lower  curve) bands
shown without offset, using Kemp et al. (\cite{Kempetal}) ephemeris as 
discussed in the text.
              }
         \label{fig4}
   \end{figure}

Background subtracted light curves were extracted in channels 5-13
(2-6 keV) and 14-23 (6-10 keV).
A Fourier analysis reveals power at the known pulse frequency,
but  the peaks are not  statistically significant. Hence, while a 
detection
of the spin period from the {\em RXTE} data alone cannot be claimed,  the
{\em RXTE} data are consistent with the known pulse period.  To fold 
light curves at the spin period, we use the Kemp et al. 
(\cite{Kempetal}) ephemeris since our X-ray ephemeris  are not 
accurate  enough to extrapolate the {\em XMM-Newton} to  {\em RXTE} 
observations.  The folded light curves in these two bands are shown in 
Fig.~\ref{fig4}. 
The modulation is present in both bands and very similar to that 
observed with great details in the EPIC data. However, given that
the source is close to the systematic limit of {\em RXTE} detection, which 
can affect the average count rate of the source especially in the harder
band, we are unable to assess on the relative pulse amplitude  in the two 
bands. We 
then limit ourselves to confirm the presence of X-ray pulse and that
the maximum of the modulation is in phase with the optical spin ephemeris
previously determined (see also Sect.\,4.2).

%
%

   \begin{figure*}[t]
   \centering
\mbox{\epsfxsize=8cm\epsfysize=8cm\epsfbox{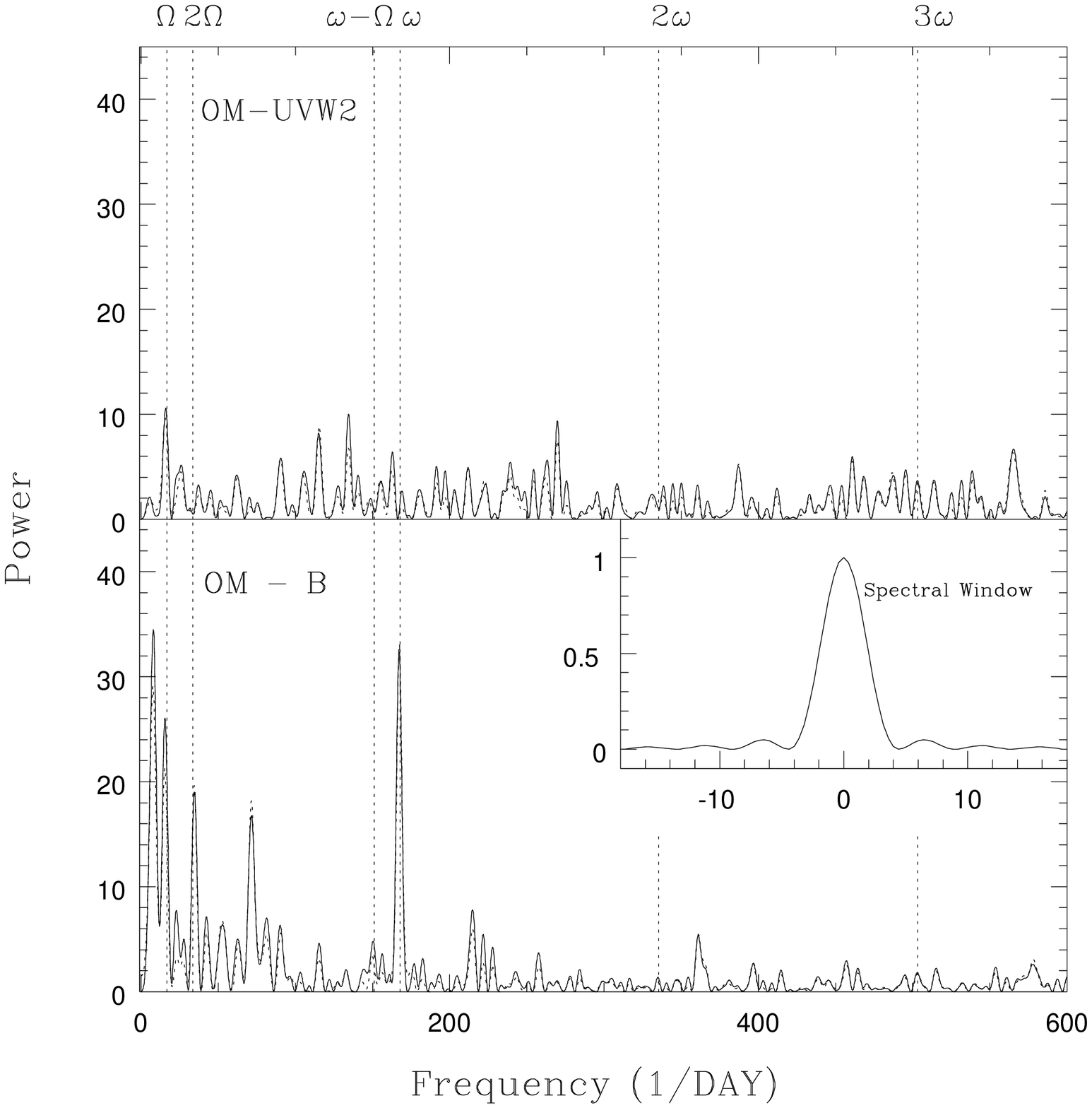}\epsfxsize=8.7cm
\epsfysize=8cm\epsfbox{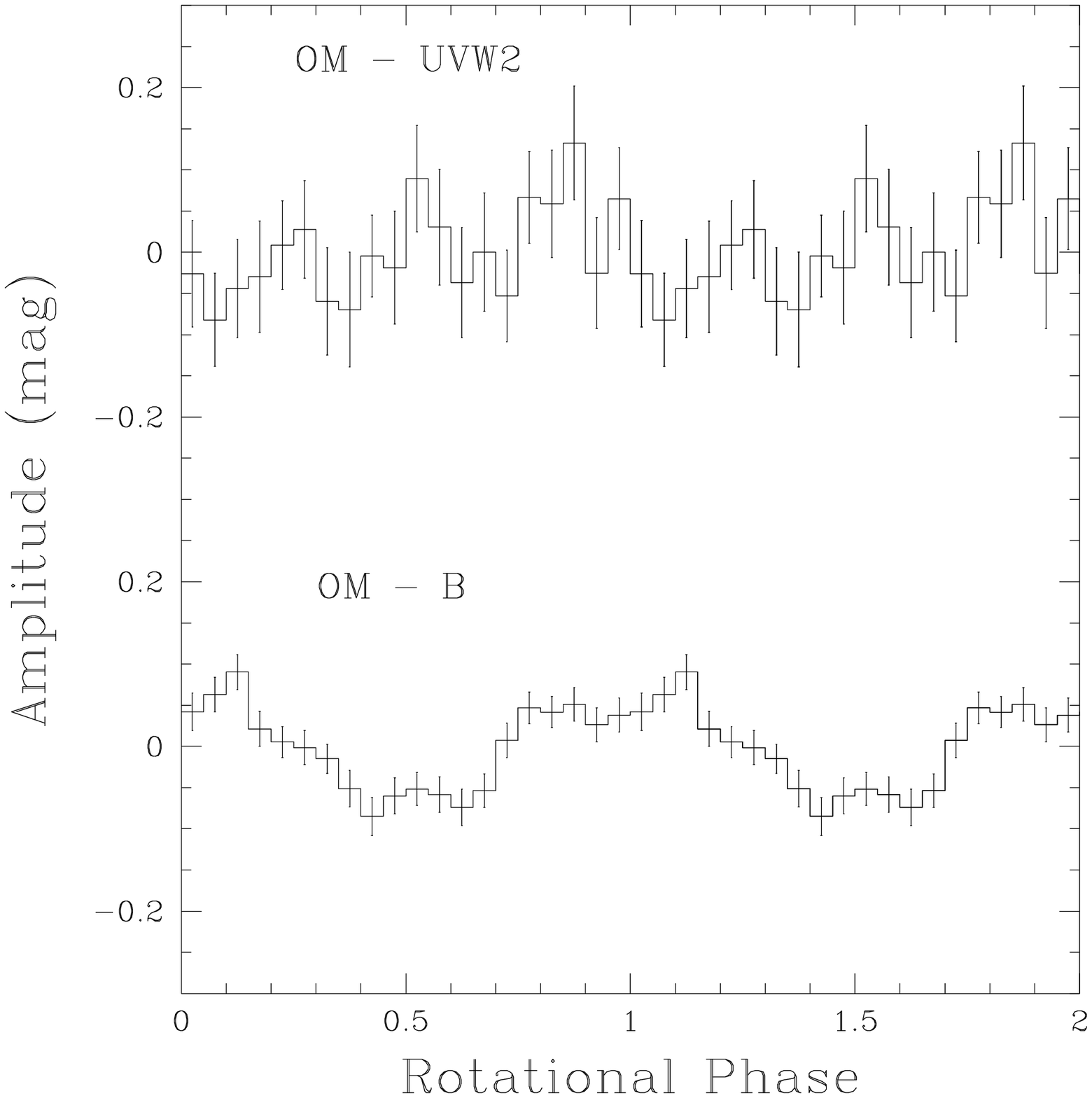}}
\caption{ {\em Left:} DFTs (solid line) of B and UVW2 time series, 
together 
with that using the CLEAN algorithm (dashed line). {\em Right:} Folded 
light curves at the 515\,s
spin period using the X-ray ephemeris quoted in the text. Ordinates are 
modulation amplitudes.}
\label{fig5}
    \end{figure*}


\section{ Optical and UV timing analysis}

\subsection{The Optical Monitor data}

\begin{figure}[b]
\centering
\mbox{\epsfxsize=8cm\epsfysize=8cm\epsfbox{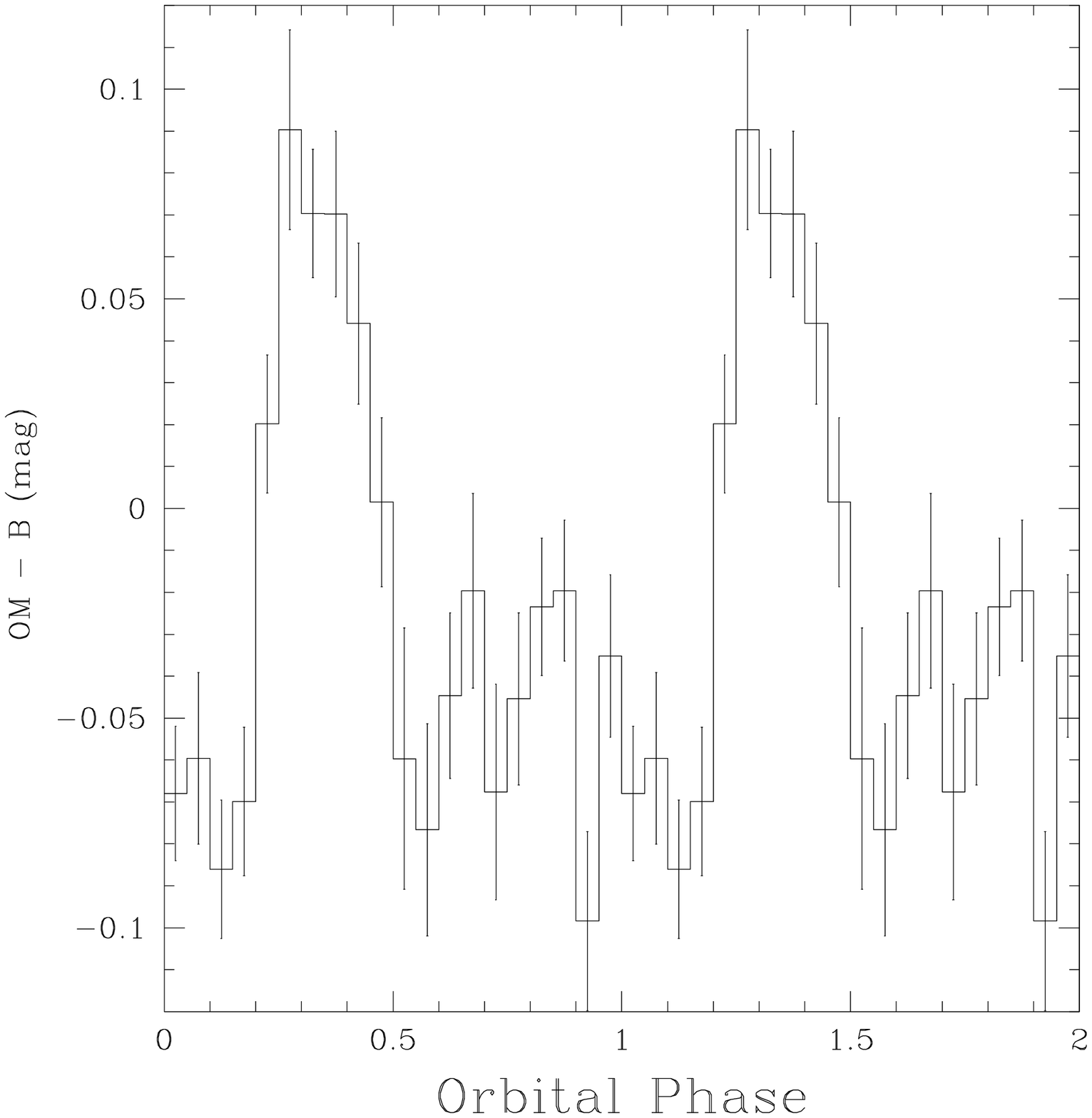}\epsfxsize=8.7cm
\epsfysize=8cm\epsfbox{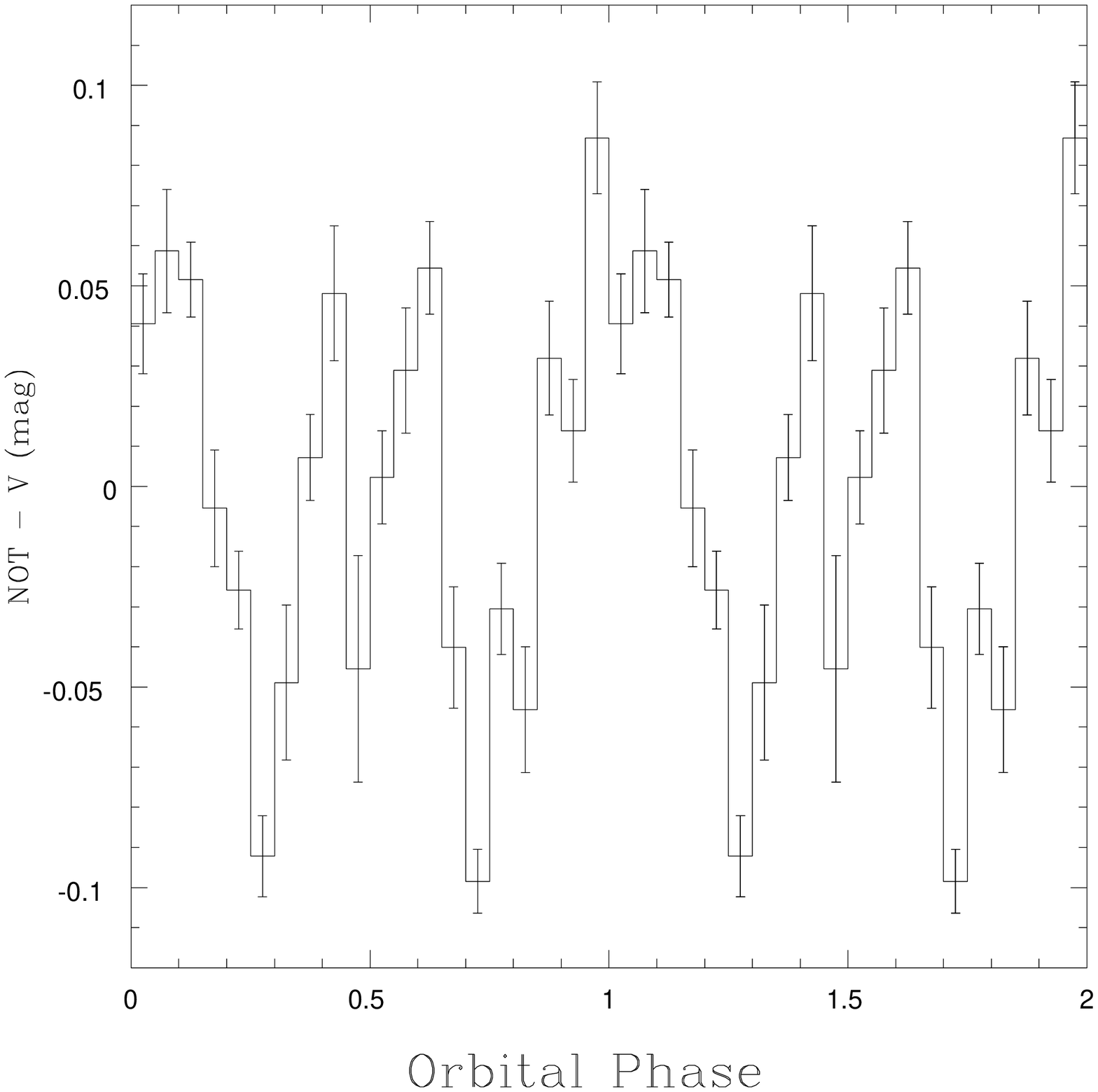}}
      \caption{The March 2003 OM B band (left) and the December 2002 {\em 
NOT} V  band (right) folded light curves at the  orbital 85.98\,min period 
after pre-whitening the data from the high frequency spin pulse. 
Phase zero has an arbitrary origin. Ordinates are modulation amplitudes.
}
         \label{fig6}
   \end{figure}

The {\em XMM-Newton\/}  OM B and UVW2 filter 10\,s light curves were 
inspected for
periodic signals. A discrete Fourier analysis (DFT) is shown in
the left panel of Fig.~\ref{fig5}, 
where the 515\,s pulse is clearly evident in the B band but not in the UV 
range. The CLEAN algorithm has also been applied thus removing most of the
windowing effects of OM data acquisition. No power is detected at the
 beat ($\omega - \Omega$) frequency   or at 
the first and second harmonics of the 515\,s signal. Power is present 
 close to the orbital (16.7\,day$^{-1}$) frequency in the B and UV 
bands although in this latter range the signal is very weak. Power is also 
detected at the first harmonic of the orbital frquency in the B band only. 
A  sinusoidal fit to 
the B band light  curve with two frequencies, accounting for the spin and 
orbital modulations,  gives $\omega$=167.06$\pm$0.26\,day$^{-1}$ and
$\Omega$=15.22$\pm$0.27\,day$^{-1}$ . The spin period agrees within 
3$\sigma$ with that determined by Kemp et al. \cite{Kempetal}, while
the orbital period is off by $\sim 5\sigma$. 
The B and UVW2 band light curves were then folded using the pulse
ephemeris derived from the EPIC data and are shown in the right panel of
Fig.~\ref{fig5}. The B band pulse fraction is 0.13$\pm$ 0.01\,mag 
consistent with previous detections and its shape is not sinusoidal 
with a broad maximum and a dip overimposed on it as well as a structured 
minimum. 
In order to inspect the orbital variability, the 
 B band light curve was pre--whitened by the 515\,s signal and then 
folded at the 85.98\,min  spectroscopic  orbital period determined  
by Kemp et al. (\cite{Kempetal}).   However, their spectroscopic 
orbital ephemeris is not accurate enough to be used for comparison between
the zero crossing of radial velocities of emission lines and the 
photometric maximum.
In Fig.~\ref{fig6} (left panel), we report the orbital light curve which 
shows a 
strongly peaked modulation with a full amplitude of 
 0.14$\pm$0.02\,mag. The strongly non-sinusoidal square shape of the 
modulation  explains the 
presence of the first harmonic of the orbital frequency.  Such 
variability 
in quiescence was not reported  previously.

%

\subsection{The UBVRI light curves}

\begin{figure*}
   \centering
\includegraphics[height=13.cm,width=7.cm,angle=-90]{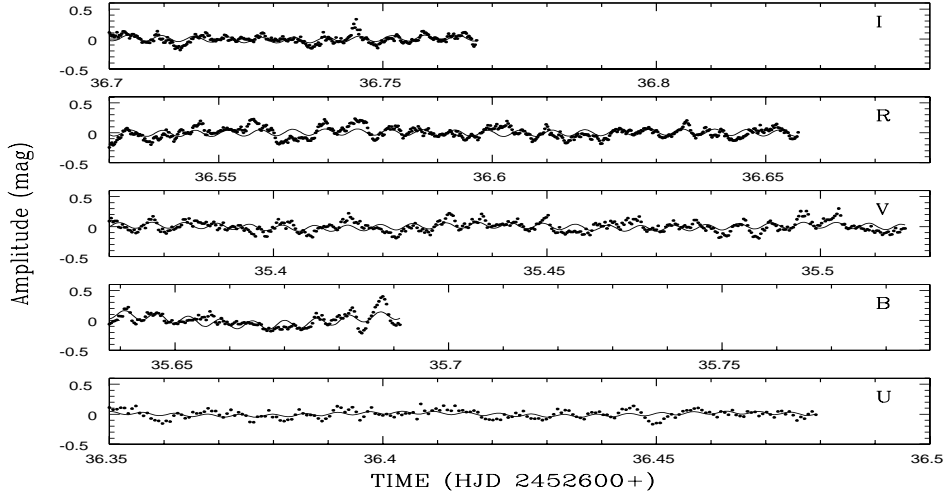}
\caption{The UVBRI {\em NOT} light curves together with their best fit
sinusoidal function with a period of 515\,s. For the B band light curve 
a polynomial function  accounting for the low 
frequency trend has been included. Ordinates are modulation 
amplitudes. 
}\label{fig7}
    \end{figure*}

\begin{figure*}
   \centering
\mbox{\epsfxsize=8cm\epsfysize=8cm\epsfbox{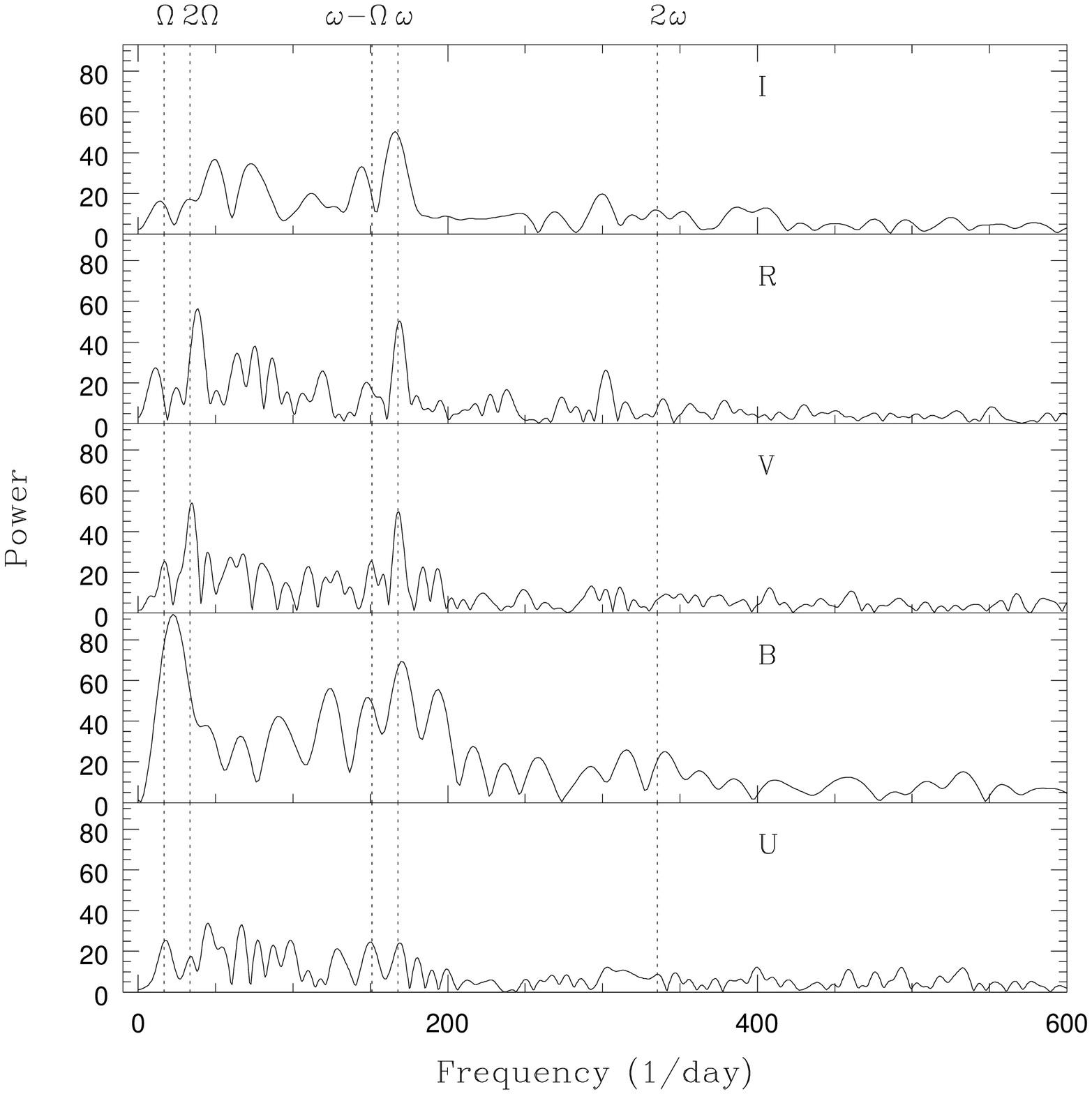}\epsfxsize=8.7cm
\epsfysize=8cm\epsfbox{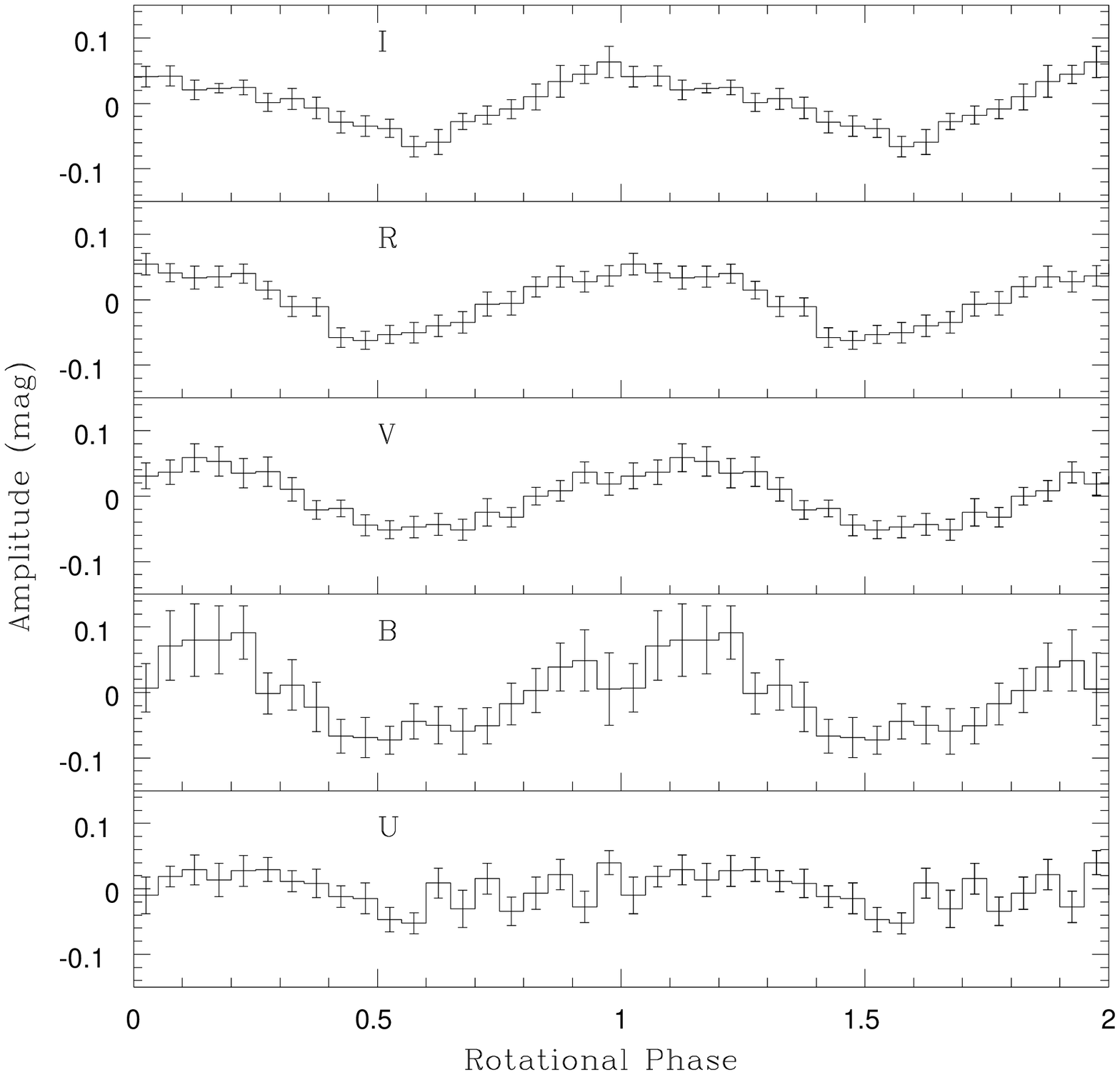}}
\caption{ {\em Left:} The DFTs of UBVRI light curves show a 
that the spin pulse is present in all bands except in the U.
 {\em Right:} Folded light curves at the
515\,s period using Kemp et al. (\cite{Kempetal}) ephemeris. These confirm 
the lack of UV pulsation but a strong optical 
modulation at longer wavelengths. Ordinates are modulation amplitudes. }
\label{fig8}
    \end{figure*}

The UBVRI light curves confirm the behaviour observed in the
OM data (Fig.~\ref{fig7}), the U band being unmodulated. Fourier analysis
of the time series is shown in the left panel of Fig.~\ref{fig8}, where 
the
515\,s pulsation is clearly detected longward  of the B band. Power 
at
the beat frequency is not statistically significant but we 
detect power close to the first harmonic of the orbital frequency in 
the V and R bands, which are the longest runs. 
Sinusoidal fits to the light curves consisting of one frequency give
$\omega$(B)=171.36$\pm$1.41\,day$^{-1}$ (this does not 
improve when
including a second frequency fixed at the orbital period),
$\omega$(V)=167.8$\pm$0.4\,day$^{-1}$, 
$\omega$(R)=168.6$\pm$0.4\,day$^{-1}$ and
$\omega$(I)=166.0$\pm$0.9\,day$^{-1}$ (see Fig.~\ref{fig7}). 
These determinations are within errors consistent with that found by Kemp 
et al. (\cite{Kempetal}). Given the low accuracy of the   frequency 
and the time of 
maximum determination, even in the longest run with the V filter, we use
the optical ephemeris by Kemp et al. (\cite{Kempetal}) to construct spin 
folded light curves. The pulses (right panel of
Fig.~\ref{fig8}) are strong at longer wavelengths but roughly similar 
(within errors) in all bands. The full amplitudes are
0.13$\pm$0.01\,mag in the B, 0.10$\pm$0.01\,mag in  
the V, 0.10$\pm$0.01\,mag in the R and  0.09$\pm$0.01\,mag in the I 
band. A comparison with the OM B band light curve shows consistency with a 
structured maximum  and a dip superposed on it. This feature 
seems to disappear moving towards the red.  As found in the simultaneous 
X-ray and optical observations of March 2003, also during the December 
2002 observations, the X-ray and optical pulses are in phase.  

 We have further inspected the data against orbital variability in the 
long runs with the V and R filters.   While we are 
unable to determine the orbital frequency  in the R band, we find 
$\rm P_{orb}=0.059\pm$ 0.003\,day in the V filter which is consistent with 
the previous optical determinations. We then detrended the V filter 
light curve from the spin modulation and  folded  the data
 at the  orbital period determined by Kemp et al. (\cite{Kempetal}). 
Fig.~\ref{fig6}, (right panel) shows a structured light curve 
with a  broad primary maximum and a double-peaked second one 
offsetted by 0.5 in phase. This explains the lack of power at the orbital 
frequency but at its first harmonic. Given that the B band 
coverage is less than one orbital period, we have compared the shape of 
the V band light curve with that observed in March 2003 in the B filter
and it appears
that the structured minimum in the latter has developed into two separate
peaks. The full amplitude of the variability being $\sim$0.15\,mag.

\section{The X-ray spectra}

\begin{figure*}
   \centering
\includegraphics[height=14.cm,width=7.cm,angle=-90]{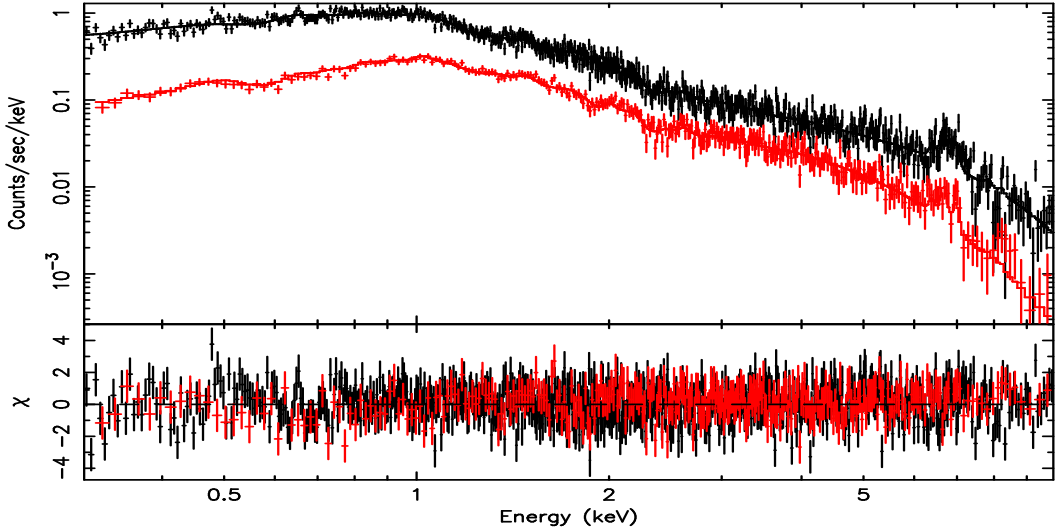}
\caption{The EPIC PN (top) and combined MOS (bottom) spectra fitted 
simultaneously with a
multi--temperature model with kT$_{\rm max}$=19\,keV and power law index
$\alpha$=0.72 absorbed by a column with density $\rm N_H=5.9\times
10^{20}$~cm$^{-2}$ plus a Gaussian centred at 6.4\,keV to account for
the fluorescent iron line.  The bottom panel shows the
residuals.
}\label{fig9}
    \end{figure*}

The average EPIC PN and combined MOS spectra were
analysed with the XSPEC package. Due to calibration accuracy issues, the 
spectra have been analysed between 0.3 and 10\,keV. An absorbed simple 
isothermal
optically thin emission model {\sc mekal }  plus a Gaussian centred
at 6.4\,keV does not provide an acceptable  fit  
($\chi^2_{\nu}=20$). The
accretion post-shock regions are expected to possess a temperature
gradient due to the cooling of the gas approaching the WD surface. We then
used a composite model consisting of a multi--temperature optically thin
plasma {\sc cemekl } with a power law temperature distribution with index
$\alpha$ and maximum temperature $\rm
kT_{max}$ and metal abundance $\rm A_Z$ in number with respect to 
solar,  plus
a zero-width (unresolved) Gaussian fixed  at 6.4\,keV and a total 
absorber with 
a hydrogen column density $\rm N_H$. This model gives a much better
agreement as shown in  Fig.~\ref{fig9}. The best fit spectral parameters,
together with the 90$\%$ confidence level for the interesting parameter,
are reported in  Table~\ref{spectra}. The hydrogen column density of the 
absorber is unusually low and consistent with the galactic
hydrogen column density in the direction of the source
$\rm N_{H,gal}=4.3\times 10^{20}$~cm$^{-2}$ (Dickey and Lockman
\cite{dl}). 
Hence, while most IPs suffer  from strong and  sometimes  
very
complex absorption (see Mukai et al. \cite{mukai}), it seems that HT Cam 
does not share this general characteristic. Also,   assuming
that the maximum temperature represents that of the shock, 
($\rm kT_{max} = kT_{shock}=3/8\,G\,M_{WD}\,\mu\,m_{H}\,R_{WD}^{-1}\,keV$, 
where $\mu$ is the mean molecular weight and $\rm m_{H}$ the mass of 
hydrogen), the inferred value is not as high as those 
determined in other systems
of this class (Beardmore et  al. \cite{beardmore00}), but it is similar to 
that determined in EX\,Hya (Fujimoto \& Ishida \cite{fuji}). The metal
abundances  are sub-solar, although caution has to be taken
since these are  relative  to Anders \& Grevesse solar abundances,
whose iron abundance is higher than others available in
XSPEC. 
The low equivalent width (EW) of the fluorescent iron line 
indicates that 
reflection is not important in this source and justifies the neglection of 
this component in the spectral fits.

The EPIC PN and combined MOS spectra were also extracted around
pulse maximum
($\phi$=0.8-1.2) and minimum ($\phi$=0.4-0.6) and fitted with the
same model, keeping fixed the
metal abundance to the value found for the average
spectrum. To fit the spectrum at minimum also the power law
index has been kept fixed to the average value. 
Leaving free  this parameter the temperature slope is less steep
($\alpha=0.51^{+0.14}_{-0.13}$), but the shock temperature is
 badly constrained ($\rm kT_{max}=30^{+20}_{-10}$\,keV,
$\chi^2_{\nu}$=0.80). As expected from the light curves and hardness
ratios at different energies, the only parameter which
changes significantly is the normalization, by a factor of 
40$\%$, the absorption remaining constant within errors. 
A further check has been done by only varying the absorption
and keeping fixed the normalization when fitting the pulse minimum 
spectrum. While absorption increases up to 
1.4$\times 10^{21}$\, cm$^{-2}$ (or 1.0$\times 10^{21}$\, cm$^{-2}$) when 
fixing the normalization at the  pulse maximum value (or at the average 
spectrum value), the fit quality is much 
worse, $\chi^2_{\nu}$=2.55 (or $\chi^2_{\nu}$=1.53). Furthermore, the EW 
of  fluorescent iron line is not 
constrained although there is 
indication that it is stronger at pulse minimum, likely due to the changes
in the continuum level. The best spectral fit parameters at these 
phases are also reported in Table~\ref{spectra}.

   \begin{table*}[t!]
      \caption{Spectral parameters as derived from  
fitting simultaneously the EPIC PN and MOS phase--averaged spectra,
those at pulse maximum and minimum  as well as from  the fits to the  
RGS1 and RGS2 averaged spectra. Quoted errors refer to 90$\%$
confidence level for
 the parameter of  interest.}
         \label{spectra}
     \centering
\begin{tabular}{c c c c c c}
            \hline \hline
            \noalign{\smallskip}
 & \multicolumn{3}{c}{EPIC} & \multicolumn{2}{c}{RGS} \cr 
 & Average  &  Pulse Maximum     &   Pulse Minimum
&  Average  & Average \cr
            \noalign{\smallskip}
            \hline
            \noalign{\smallskip}
$\rm F_{2-10}^{a}$ & 3.36$\pm$0.02 & 3.84$\pm$0.05 &
2.65$\pm$0.06 & & \cr
            \noalign{\smallskip}
            \hline
            \noalign{\smallskip}
$\rm N_H^{b}$    & 5.9$^{+0.7}_{-0.4}$  & 5.2$\pm$0.6 &
6.7$^{+0.9}_{-1.0}$  & 5.9 (fixed) & 11.0$^{+3.0}_{-2.0}$ \cr
$\rm kT_{max}[keV]$ & 19$\pm$3   & 19$^{+5}_{-3}$ &
18$^{+3}_{-2}$ & 19 (fixed) & 19 (fixed)  \cr
$\alpha$ & 0.72$^{+0.10}_{-0.08}$ & 0.74$\pm$0.1 & 0.72
(fixed) & 0.72 (fixed) & 0.72 (fixed) \cr
$\rm A_Z^{c}$ & 0.59$\pm$0.10 & 0.59 (fixed)  & 0.59 (fixed) & 
0.59 (fixed) & 0.59 (fixed) \cr
$\rm C_{nor.}^{d}$ $[10^{-3}$] & 5.1$^{+0.4}_{-0.3}$ &6.0$^{+0.6}_{-0.7}$
& 3.9$\pm$0.1 & 3.9$\pm$0.2 & 4.6$^{+0.4}_{-0.5}$ \cr
$\rm E.W.^{e}$ & 50$^{+34}_{-40}$ & 27($<80$) & 100($<$150) & & \cr
                 \noalign{\smallskip}
                 \hline
            \noalign{\smallskip}
$\chi^2_{\nu}$ ($\chi^2$/d.o.f.) & 1.02 (1082/1058) & 0.91
(561/615) &0.81 (279/344) & 0.96 (200/209) & 0.92 (191/208) \cr
                 \noalign{\smallskip}
            \hline
\end{tabular}

~\par
\begin{flushleft}
$^a$: Flux in units of $10^{-12}\,\ecs$ in the 2--10\,keV band.\par
$^b$: Column density of the absorber in units of $10^{20}$~cm$^{-2}$.\par
$^c$: Metal abundance in units of the cosmic value (Anders \& Grevesse
\cite{andersgrevesse}).\par
$^d$: Normalization constant of {\sc cemekl } model.\par
$^e$: Equivalent width of the 6.4 keV fluorescent iron line in units 
of eV.\par
\end{flushleft}
\end{table*}

\begin{figure}
\centering
\includegraphics[width=8cm]{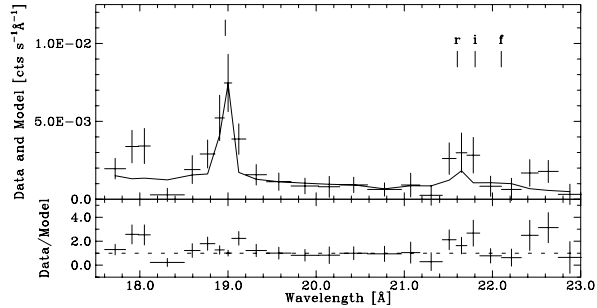}
      \caption{The region including the Oxygen lines of the RGS
spectra, together with the EPIC best fit model. The rest wavelengths of 
the lines are also
marked. The positions for the forbidden (f), intercombination
(i) and resonance (r) lines are displayed. 
              }
         \label{fig10}
   \end{figure}

In the RGS spectra, we detect the {\sc O\,VIII} (19.1\,$\AA$) 
Ly$_{\alpha}$ line with high degree of confidence,
including the line width ($\sigma \sim 1100\pm 600 \rm
km\,s^{-1}$).  The {\sc O\,VII} (21.9\,$\AA$) He-like triplet is also
detected.  The Oxygen region of the RGS spectra are shown in
Fig.~\ref{fig10}. The detected  {\sc O\,VII} line is
likely to be a blend of the resonant (r) and
intercombination  (i) lines,
whereas there is no evidence for the forbidden (f) line.  We
also detect a feature around 15.1\,$\AA$, which is likely to be a
blend including the Fe\,{\sc XVII} 15.01\,$\AA$ line.
 There is an excess of counts around 17\,$\AA$ that is consistent 
with the Fe\,{\sc XVII} lines expected there, given the presence of 
the 15.1\,$\AA$  blend.
However, the 17\,$\AA$  feature is too weak  to provide reliable
numbers.
The measured line parameters are reported in Table~\ref{rgspar}.

   \begin{table}
      \caption{Line parameters as measured in the RGS spectra of
HT\,Cam.}
         \label{rgspar}
     \centering
\begin{tabular}{c c c c }
            \hline \hline
            \noalign{\smallskip}
Line & $\rm E_C$ & $\sigma$       &  Flux \cr
            \noalign{\smallskip}
     &   (keV) & (eV) & ph\,cm$^{-2}$\,s$^{-1}$\cr 
            \noalign{\smallskip}
            \hline
            \noalign{\smallskip}
Blend  &  0.8213$_{-0.0044}^{+0.0046}$ & 6.2$_{-3.4}^{+3.8}$ & 
2.0$\pm 0.9 \times 10^{-5}$ \cr
{\sc O\,VIII} & 0.6536$_{-0.0009}^{+0.0011}$ &
2.4$_{-1.3}^{+1.4}$ & 3.3$\pm 0.9\times 10^{-5}$ \cr
{\sc O\,VII} &0.5723$_{-0.0030}^{+0.0031}$ & 2.8($<5.8$) &
2.1$_{-1.3}^{+1.5}\times 10^{-5}$ \cr
            \noalign{\smallskip}
            \hline
\end{tabular}
\end{table}

%

The same multi--temperature model used for EPIC spectral fits
provides an  adequate description of the RGS spectra.  We obtain
$\chi^2_{\nu}$= 0.97 by just adjusting the normalization, with 
all other parameters ($\rm T_{max}$, $\alpha$, $\rm
A_Z$ and $\rm N_H$) fixed to the EPIC spectral fit values).  The
resulting normalization
for the RGS spectra is somewhat smaller than the best-fit EPIC
value.
We obtain a better fit ($\chi^2_{\nu}$=0.92) allowing absorption 
to vary; the best-fit $\rm N_H$ value is higher (1.1$\times
10^{21}\, \rm cm^{-2}$) than in the EPIC spectral fit,
while the normalization is in better agreement.  Because of the
limited statistical quality, potential systematic errors in the background
subtraction, and/or spectral complexity (HT\,Cam may have low
level partial covering absorber, as is seen in EX Hya (Cordova et
al. \cite{cordova})), we do not consider these discrepancies to
be serious.  Instead, we take the success of the {\sc cemekl} fit
as confirming the multi-temperature nature of the X-ray emitter
in HT\,Cam.

\section{Discussion}

Our X-ray observations of HT\,Cam reveal a strong (40$\%$) pulsation at 
the 515\,s optical period, which  unambiguously assigns this CV to the 
IP class of magnetic systems. It is then the third confirmed IP, 
 along
with V\,1025 Cen and EX\,Hya, below the 2-3\,hr orbital period gap.

\subsection{The pulsator in HT\,Cam}

The X-ray power spectrum does not reveal further periodic signals, 
such as sidebands, which indicates  
that the X-ray period is indeed the rotational period of the WD and 
that accretion occurs via a disc (Norton et al 
\cite{nortonetal}).
The X-ray pulses are quasi-sinusoidal as  in  most IPs, but 
 different  from 
the majority of these systems they are not energy dependent. In the 
classical accretion scenario, the  accreting material falls onto the WD 
poles from the inner edge of the  truncated disc at the magnetospheric 
radius and flows along the magnetic field lines  in an arc-shaped curtain 
(Rosen et al. \cite{rosen}). In this model, the optical depth of the 
infall gas in the 
curtain is larger along the field lines rather than perpendicularly and hence 
photoelectric absorption is larger when the curtain points towards the
observer. As for the short orbital period IPs, 
EX\,Hya (Rosen et al. \cite{rosen}) and  V\,1025 Cen 
(Hellier et al.  \cite{hellier98}), the mechanisms causing the pulsations 
are a combination of photo-electric absorption in the curtain and 
occultation.  
In HT\,Cam, instead there is no evidence for phase--dependent
absorption. This suggests that the rotational pulses are only due to 
aspect changes of the  X-ray 
emitting region, i.e. occultation of this region as the WD rotates. The 
changes by 40$\%$ of the normalization implies that the emission measure 
of the X-ray emitting source  changes from 4.7$\times 10^{53}\,\rm 
d_{100}^2\,cm^{-3}$ at pulse minimum to 7.2$\times  10^{53}\,\rm 
d_{100}^2\,cm^{-3}$ at pulse maximum,  where  $d_{100}$ is 
the distance in units of 100\,pc.
Hence, although we cannot exclude that a secondary pole is also visible, 
it is reasonable to assume that the main accreting pole does not 
completely disappear behind the WD limb. 

 The optical pulses are in phase with the X-ray modulation, similarly to 
most IPs. This implies that the optical pulsation originates in the 
magnetically confined accretion flow.  
The pulse amplitudes in our colour photometry are quite similar  
in all bands ($\sim$10$\%$) except in the UV where both far-UV and near-UV 
ranges are unmodulated. This is an unusual behaviour, since the UV flux 
is  known to be modulated at the rotational period of the WD in other 
IPs, like AE\,Aqr (Eracleous \& Horne \cite{eracleoushorne}), YY\,Dra 
(Haswell et al. \cite{haswelletal}), FO\,Aqr (de Martino et al. 
\cite{demartinoetal}), EX\,Hya (Eisenbart et 
al. \cite{eisenbartetal}). In these systems the UV pulses originate in 
the  heated polar region of the WD atmosphere. Hence, the lack of UV 
modulation in HT\,Cam suggests that heating of the WD poles is 
negligible. Also, it 
is very unlikely that the  reprocessing of hard X-rays peaks out of the 
UV window, since a soft X-ray component is not observed in  the 
X-rays.  Furthermore, the optical pulsations in IPs are typically colour 
dependent (Welsh \& Martell \cite{welsh}, de Martino  et al. 
\cite{demartinoetal}, Eisenbart et al. \cite{eisenbartetal}) and  
the pulse spectrum is generally  described by a blackbody function. In 
HT\,Cam the lack of colour dependence, together 
with the absence of UV pulses is difficult to reconcile 
with  optically thick emission. An  alternative  possibility is 
that
the optical spin pulses are due to aspect angle changes of a 
 region in the accretion flow which is optically thin but 
cool enough to not affect the Balmer jump. 
The low hydrogen column  density found in the X-rays might corroborate 
this idea, but   further photometry extending to the IR and 
phase--resolved optical spectrophotometry   
 are  required  to clarify this puzzle. 

\noindent We also detect an orbital strongly peaked modulation in 
the 
B band photometry from the OM instrument (March 2003) with full amplitude 
of  0.14\,mag. This is also detected in the December 2002 
{\em NOT} run but the modulation  appears to have 
changed  into a complex double structured shape.
 This variability during quiescence was not reported previously, 
but only when HT\,Cam was in outburst. We are confident that HT\,Cam 
was in quiescence during both optical observations in December 2002 and 
March 2003. It might be possible that the orbital modulation was 
 over-looked in previous optical photometric data sets.

\subsection{The post--shock region}

Both EPIC and RGS spectra reveal that the X-ray post--shock emitting 
region has a multi--temperature structure. This is expected by the 
standard accretion shock model, where matter falls onto the WD almost 
radially at 
free-fall velocities thus producing a stand-off shock below which gas
cools via thermal bremsstrahlung as it descends and settles on the WD 
surface (Aizu, \cite{aizu}).
From the ratio of Oxygen lines, which particularly map the 
low temperatures, and the maximum temperature determined from the 
spectral fits, the  plasma temperature extends from 
kT$_{\rm min}=0.3^{+0.2}_{-0.1}$\,keV 
to  kT$_{\rm max}\sim$19$\pm$ 3\,keV.
This range of temperature is, within errors, close to that found in 
EX\,Hya   (Fujimoto \& Ishida \cite{fuji}). To infer the post--shock 
densities,  the standard 
diagnostic of high-temperature plamas is the ratio of forbidden to 
intercombination lines of He-like ions (R=f/i). This ratio decreases with 
increasing densities. In HT\,Cam the forbidden line of {\sc O\,VII} is not 
detected,  suggesting a high density limit
(Porquet \& Dubau \cite{porquet}).  However, the quality of 
the RGS spectrum can provide only a 1$\sigma$ lower 
limit  on the electron density: $n_e>5\times 10^{12}$\,cm$^{-3}$.
We also note that the emission measure derived above is remarkably 
in agreement with that expected from  post-shock regions. For a typical 
polar cap $\rm \sim 10^{16}\,cm^2$ and shock height of $\rm 10^8\,cm$, the 
emission measure  gives a gas density of 
 $n \sim 8\times10^{14}\,d_{100}\,\rm cm^{-3}$, which is consistent 
with  the high density limit  estimated  from the RGS data. 
However, as stressed by Porquet et al. (\cite{porquet01}) and Mauche et 
al. (\cite{mauche}) a strong UV radiation field, as present in    
CVs, can mimic a high density regime, implying that   
the upper level of the forbidden line can be significantly depopulated via 
photo-excitation to the upper levels of intercombination lines. 
As for the {\sc O\,VII} line, a blackbody radiation with temperature of 
$>$30000\,K can produce a strong decrease in the intensity of the 
forbidden line and an increase of the intercombination line.

  Furthermore, expected typical
post--shock  velocities in magnetic CVs are $\rm v\leq v_{shock}=v_{ff}/4 
\sim$\,900\,km\,s$^{-1}$, 
where $\rm v_{ff}=\rm (2G\,M_{WD}/R)^{1/2}$ is the free-fall veolcity in 
the  pre--shock flow. The optical spectrum of HT\,Cam (Tovmassian 
et al. \cite{Tovmassianetal}, Kemp et al. \cite{Kempetal}) shows broad 
emission lines (FWHM$\sim$3000-4000\,km\,$^{-1}$), while 
the width of the X-ray Oxygen lines, though not well constrained, are 
narrower ($\sigma\sim 1000\pm 600$\,km\,s$^{-1}$). Hence, the X-ray lines 
are broadly compatible with post--shock velocities.   

 Using the maximum temperature derived from the X-ray spectral fits 
(19$\pm$3\,keV) as the shock temperature and adopting the WD mass-radius 
relation of Nauenberg 
(\cite{nauenberg}), we derive 
a WD mass  $\rm M_{WD}=0.55 \pm 0.09\,M_{\odot}$. This value falls
in the low mass range of CV masses and it is similar to that determined 
for EX\,Hya (Fujimoto \& Ishida \cite{fuji}, Cropper et al. 
\cite{cropper}).

The weak fluorescent "neutral" Fe line at 6.4\,keV indicates that 
reflection from cold material is not important in HT\,Cam. The 
origin of the fluorescent line in magnetic CVs is likely reflection of 
hard X-rays at the WD surface (Beardmore et al. 
\cite{beardmore00}, Matt et al. \cite{matt}, de Martino et al. 
\cite{demartino01}, \cite{demartino04}), although the 
pre--shock material can be also an additional source
(Ezuka \& Ishida \cite{ezukaishida}). However, the 90$\%$ confidence level 
lower limit to the EW of the 6.4\,keV line is too large to be 
produced by a pre-shock gas with hydrogen column density $<10^{21}$\,cm$^{-2}$, 
implying that this line originates by reflection at the WD surface.

 \subsection{HT\,Cam:  a low mass transfer rate system}

The hydrogen column density of the absorbing material above the shock is 
found to be lower by two orders of  magnitude than typical values of 
IPs (Ezuka \& Ishida \cite{ezukaishida}). Here we note that also the other 
two 
IPs below the gap, EX\,Hya (Mukai et al.  \cite{mukai03}) and 
V1025\,Cen  (Hellier et al. \cite{hellier98}) are characterized by low 
absorbing columns, although these two possess phase--dependent complex 
partial absorbing material.  Hence,  the lack of
a high density   absorber above the source suggests that the mass 
accretion rate is low. The large EWs 
of  optical Balmer lines and the weak  He\,II emission feature 
(Tovmassian et al. \cite{Tovmassianetal}) are now understandable in this 
context. We use the unabsorbed bolometric flux derived from spectral fits  
which gives:  $\rm L_{bol}=9.0\times10^{30}\,d_{100}^2$\,\es.  
Equating this to the accretion luminosity $\rm L_{accr.}= 
G\,\dot M_{\odot}\,M_{WD}/R_{WD}$ and assuming the derived WD mass $\rm 
M_{WD}$=0.55\,$\rm M_{\odot}$, we derive $\rm \dot M = 2.4\times 
10^{-11}\,d^2_{100}\,M_{\odot}\,yr^{-1}$.  The predicted accretion rate
for a system with $\rm P_{orb}$=86\,min is $\rm \dot M \sim 2\times 
10^{-11}\,M_{\odot}\,yr^{-1}$. Hence, unless the distance is as large as 
that estimated by Tovmassian et al. (\cite{Tovmassianetal}) (d=400\,pc), 
the accretion rate is consistent  with that predicted by 
gravitational 
radiation and similar to that recently found in EX\,Hya (Beuermann et al.
\cite{beuermann}).
We note that, Tovmassian et al. (\cite{Tovmassianetal}) estimate the
distance for  HT\,Cam from K-band flux assuming a M8\,V secondary star with 
radius similar to normal field dwarf stars. Probably this  is  
unlikely (see also below), but also because the mass accretion rate 
would be  too high for transfer rates expected for short orbital 
period systems (Patterson \cite{patterson98}).

Hence, HT\,Cam shares the low accretion rates with the other two IPs 
below the  period gap, EX\,Hya and V1025\,Cen, but differently from these 
weakly de-synchronized systems, the spin-to-orbital period ratio is close 
to 0.1. This value  is the one expected for systems spinning at 
equilibrium by accreting the  specific angular momentum of the secondary star (King \& Lasota 
\cite{kinglasota}). Indeed, Kemp et al. (\cite{Kempetal}) find that 
the spin period derivative  in HT\,Cam is small. Thus HT\,Cam is located 
at the border line  in the spin-orbital period plane above which  
accretion does not occur predominantly via a disc. However, for the 
three short-period systems   indications of  disc accretion are 
found. 
Assuming spin equilibrium and disc accretion, a value of the magnetic 
moment $\mu$ of the WD can be obtained: $\rm \omega_s = 
0.041\,P_{1000}^{-1}\,\dot M_{17}^{-3/7}\,M_{WD}^{-5/7}\,\mu_{32}^{6/7} $
where $\rm \omega_s$=0.6 (Patterson \cite{patterson}) is the fastness 
parameter which is valid for rapid rotators like HT\,Cam, $\rm P_{1000}$ 
is the spin period in units of 1000\,s, $\dot M_{17}$ is the accretion 
rate in units of $\rm 10^{17}\,g\,s^{-1}$ and $\mu_{32}$ is the WD 
magnetic moment in units of $\rm 10^{32}\,G\,cm^3$. Using the values 
inferred for the WD mass and the mass accretion rate,
the magnetic moment can be estimated: $\mu_{32}=0.79\,d_{100}$. 
Hence, the  fast rotator and low 
accretor in  HT\,Cam  appears to possess a weak magnetic field  
comparable to YY\,Dra, which is also rotating and 
accreting at  a similar 
rate, although it is likely far from equilibrium (Patterson 
\cite{patterson}). Hence, there is a large difference with 
EX\,Hya and V1025 Cen as they possess slowly rotating WDs and
stronger magnetic fields. We here note that
Norton et al. (\cite{norton04}) derive $\mu_{32}$=2.7 for HT\,Cam from
their model of  magnetic accretion, but they assume a mass ratio $\rm 
q=M_{sec}/M_{WD}$=0.5 and a WD mass of 0.7\,$\rm M_{\odot}$.
Since the magnetic moment controls the 
spin rate (Patterson \cite{patterson}), it might be reasonable to suppose  
that HT\,Cam will  spin up and spin-down around equilibrium 
value and, 
given  its short orbital period, probably it will never synchronize.

With the derived WD mass it is possible to obtain information on the 
mass of the donor star. From the orbital 
period and the WD orbital velocity  $\rm V_{WD}$ the mass function can
be computed as: $\rm f(m) = (M_{sec}
sin\,i)^{3}/(M_{WD} + M_{sec})^{2}= (V_{WD})^{3}P_{orb}/2\pi G$,
where $\rm M_{sec}$ is the mass of the secondary star and {\em i} the
inclination of the orbit to the
line of sight. For HT Cam, assuming that the emission line velocities
K = $(69\pm4)$ km\,s$^{-1}$ (Kemp et al. 2002)   trace the WD orbital 
motion (this value is also similar to that in  EX\,Hya (Eisenbart
\cite{eisenbartetal})), the mass function is
found to be: $\rm f(m) = (2.03\,\pm0.35) \times 10^{-3} M_{\odot}$. 
Assuming our  derived WD mass:
$\rm M_{sec}=0.09M_{\odot}$ for $i=90^{\circ}$, $\rm M_{sec}$=0.14 for
$i=45^{\circ}$ and 0.21$\rm M_{\odot}$ for $i=30^{\circ}$. 
Since no X-ray eclipses are observed, $i<75^{\circ}$. 
Also, from the presence of 
double-peaked emission lines in the optical spectrum with 
half-peak separation of 
250\,km\,s$^{-1}$ (Kemp et al. \cite{Kempetal}), it is possible to 
further constrain the system inclination.  Since these originate in the
accretion disc (see Warner \cite{warner95}), the velocity 
separation of the peaks is: $\rm V_{peak}=1154\, 
(M_{WD}/M_{\odot})^{1/2}\, 
(f_{Lobe}\,R_{Lobe}/10^{10}\,cm)^{-1/2}\, sin\,i$, 
where $\rm f_{Lobe}$ is the filling fraction of the primary star Lobe, 
usually taken as 
90$\%$.  Adopting $\rm M_{WD}$=0.55\,M$_{\odot}$ and  $\rm 
M_{sec}$=0.1-0.3\,$\rm M_{\odot}$,  then $\rm R_{Lobe}=1.82-2.05\times 
10^{10}\,cm$ and  hence $\rm V_{peak}$ =630-660\, sin\,i \,km\,s$^{-1}$. 
From  the observed value of peak separation, this yelds to  $i\sim 
25^{\circ}$.
 For comparison, EX\,Hya ($i=78^{\circ}$, Hellier et al. 
\cite{hellier87}) 
has peak separations of 
1300\,km\,s$^{-1}$. 
Hence  the secondary mass seems to be further restricted towards high 
values. For an orbit with $\rm P_{orb}$=86 min, the assumption of 
a Roche lobe filling companion
(Paczynski \cite{paczynski}), leads to a further mass-radius constrain:  
$\rm R_L$ = 0.0152 ($\rm M_{sec}$/$M_{\odot}$)$^{1/3}$\,($\rm P_{orb}$/1
min)$^{2/3}$\,R$_{\odot}$. 
A comparison with the mass-radius 
relation for main-sequenc stars (Baraffe et al. \cite{baraffe}) indicates 
that only a very low mass companion $\rm M_{sec} \leq
0.13M_{\odot}$ will fill its lobe, but for larger values as it seems to be 
suggested by the very low inclination, 
a main-sequence star will overfill its lobe. In this latter case the 
companion in HT\,Cam is significantly overmassive for its size or 
alternatively its radius is much smaller than that of a main-sequence star 
of comparable mass. 
Although the nature of the companion star is beyond the scope 
of the 
present work,  we note here  that HT\,Cam shows intriguing 
characteristics 
which new optical spectrophotometry should address.

\section{Conclusions}

We have presented the first X-ray pointed observations of HT\,Cam 
with {\em XMM-Newton} and  {\em RXTE} satellites 
together with simultaneous optical/UV photometry which 
firmly  confirm that this CV is an IP member. For its short 86\,min 
orbital  period it is then the third IP  below the 
2-3\,hr period gap.
Here we summarize the main results:

\begin{itemize}

\item  The X-ray emission is  dominated by the 515\,s optical 
modulation with no sign of signals at sideband frequencies. This indicates 
that the X-ray period is the spin period of the WD and that HT\,Cam mostly 
accretes via a disc.\\

\item The X-ray pulse is not energy dependent indicating that the 
mechanism responsible for spin pulsation is occultation of the emitting 
region while the WD rotates. \\

\item The X-ray spectra show that the post-shock region is 
multi--temperature with a maximum temperature of $\sim$ 20\,keV and that 
the photo-electric absorption above the shock is very low and constant 
throughout the spin cycle. \\

\item Oxygen lines detected in the  low-energy portion of the X-ray 
spectrum  indicate that the post-shock region is in the high density 
limit ($n_e >5\times 10^{12}\,\rm cm^{-3}$), athough  we cannot exclude 
that additional/alternative contribution from a strong UV emission is 
responsible for the lack of the forbidden line.\\

\item From the shock temperature, we estimate the WD mass: 
$\sim$ 0.6\,$\rm M_{\odot}$, which falls into the low value range of WD 
masses in CVs. \\

\item  The optical pulsation shows similar fractional amplitudes
 longward of  the blue, but surprisingly the UV flux is not 
modulated, indicating that the X-ray heating of the WD poles is not 
important in this system.\\

\item The mass accretion rate is low: 2.4$\rm \times 
10^{-11}\,d_{100}^2\,M_{\odot}\,yr^{-1}$. For d=100\,pc 
this is close to the expected value for mass transfer rates driven by
gravitational radiation.\\

\item The ratio of spin-orbit periods is close to the value predicted for 
a WD spinning at its equilibrium value and we estimate a 
magnetic moment: $\rm log\, \mu = 31.9 + log\,d_{100}\, G\,cm^3$.\\

\end{itemize}

 \begin{acknowledgements}
      Part of this work was supported by the Italian Minister of Research
(MURST).The authors wish to thank the {\em XMM-Newton} staff at VILSPA and
in particular Drs. N. Schartel, P. Rodriguez-Pascual and M. Santos-Leo for
their useful suggestions on data analysis. Some of the data presented 
here have 
been taken using ALFOSC, which is owned by the Instituto de Astrofisica de 
Andalucia (IAA) and operated at the Nordic Optical Telescope under 
agreement between IAA and the NBIfAFG of the Astronomical Observatory of 
Copenhagen.

 \end{acknowledgements}

\end{document}